\documentclass[aps,prd,twocolumn,preprintnumbers,
superscriptaddress,
floatfix]{revtex4}
\usepackage{amsmath}
\usepackage{accents}
\usepackage{epsfig}
\usepackage{graphics}
\newcommand{\eeq}{\end{equation}}
\newcommand{\beq}{\begin{equation}}
\newcommand{\ba}{\begin{array}}
\newcommand{\ea}{\end{array}}
\newcommand{\bea}{\begin{eqnarray}}
\newcommand{\eea}{\end{eqnarray}}
\newcommand{\vev}[1]{\langle #1\rangle}

\newcommand{\eps}{\epsilon}

\newcommand{\al}{\alpha}

\newcommand{\mrm}[1]{\mbox{\rm #1}}

\newcommand{\effn}[1]{\accentset{(#1)}}

\newcommand{\rd}{rd.\ }

\begin{document}

\title{Renormalization effects on neutrino masses and mixing
in a string-inspired SU$(4)\times$SU$(2)_L\times$SU$(2)_R
\times$U$(1)_X$ model}

\author{T.\ Dent}
\affiliation{University of Heidelberg, Philosophenweg 16, D-69120 Germany}
\author{G.K.\ Leontaris}
\author{A.\ Psallidas}
\author{J.\ Rizos}
\affiliation{Theoretical Physics Division, Ioannina University, GR-45110 Ioannina, Greece}

\date{\today}

\begin{abstract} \noindent
 We discuss renormalization effects on neutrino masses and mixing
 angles in a supersymmetric string-inspired SU$(4)\times$SU$(2)_L\times
 $SU$(2)_R\times$U$(1)_X$ model, with matter in fundamental
 and antisymmetric tensor representations and singlet Higgs fields
 charged under the anomalous $U(1)_X$ family symmetry.
 The quark, lepton and neutrino Yukawa
 matrices are distinguished by different Clebsch-Gordan coefficients.
 The presence of a second $U(1)_X$ breaking singlet with
 fractional charge allows a more realistic,
 hierarchical light neutrino mass spectrum with
 bi-large mixing. By numerical investigation we find a region
 in the model parameter space where the neutrino mass-squared
 differences and mixing angles at low energy are
 consistent with experimental data.
\end{abstract}


\maketitle

%\section{Introduction} \label{sec:intro}

\noindent The experimentally measured values of gauge coupling
constants $\alpha_{3},\alpha_{em}$, and the weak mixing angle
$\sin^2\theta_W$ are correctly predicted in the Minimal
Supersymmetric Standard Model (MSSM), assuming a unification scale
of the order $10^{16}$ GeV. Moreover, the existing data from
neutrino oscillation experiments provide an important clue to
physics beyond the successful Standard Model (SM) and MSSM.

Experimental data on atmospheric and solar neutrino oscillations
\cite{Bahcall:2004ut} imply tiny but non-zero neutrino
mass-squared differences $\Delta\,m^2_{\nu_{ij}}$. The negligible
size of the neutrino masses, as compared to those of quarks and
charged leptons, might suggest that a theory beyond MSSM should
incorporate the right-handed neutrinos and an appropriate
(see-saw) mechanism to suppress adequately the neutrino masses.
Moreover, the observed large neutrino mixing angles
$\theta_{\nu_{12}},\theta_{\nu_{23}}$ present challenges for
additional symmetries and a unified framework in which neutrinos
and quarks form part of same multiplet. Examples of higher
symmetries including the SM gauge
 group and incorporating the right-handed neutrino in the
 fermion spectrum, are the partially unified Pati--Salam model, based on
 $SU(4)\times SU(2)_L\times SU(2)_R$ \cite{Pati:1974yy}, and the
fully unified $SO(10)$. When embedded into perturbative string or
D-brane models, these may be extended by additional abelian or
discrete fermion family symmetries.
Thus fermion masses and mixing angles can be
compared to the predictions of various types of models with full or
partial gauge unification and flavor symmetries.

Recently some models have been proposed to explain the presence
of large mixing angles in the neutrino sector, in contrast to the
smaller quark mixings.
%has been discussed in a number of recent works
%and a variety of possible solutions have been proposed.
For example, the mixing angle $\theta_{\nu_{12}}$ and
the Cabibbo mixing $\theta_C$ could satisfy the so called Quark-Lepton
Complementarity (QLC) relation $\theta_{\nu_{12}}+\theta_C\approx
\frac{\pi}4$~\cite{Petcov:1993rk}. It has been
shown~\cite{Minakata:2004xt} that this relation can be
reproduced if some symmetry exists among quarks and leptons.
Attempts to realize QLC in the context of models unifying quarks and
leptons such as Pati-Salam have been made~\cite{Antusch:2005ca}.

As another possibility, the symmetry $L_e-L_{\mu}-L_{\tau}$
implies an inverted neutrino mass hierarchy and bimaximal mixing
$\theta_{\nu_{12}}=\theta_{\nu_{23}}=\frac{\pi}4$, with
$\theta_{\nu_{13}}=0$~\cite{lelmlt}. This symmetry alone does
not give a consistent description of current experimental data,
but
%it could be a starting point for an accurate prediction if
additional corrections and renormalization effects have still
to be taken into account. It has been shown~\cite{Leontaris:2004rd}
in the context of MSSM extended by a spontaneously broken $U(1)_X$
factor,
%and a singlet Higgs acquiring no-zero vacuum expectation value (vev),
that the neutrino sector respects an
$L_e-L_{\mu}-L_{\tau}$ symmetry. Small corrections from other
singlet vevs, which are usually present in a string spectrum, can
lead to a soft breaking of this symmetry and describe accurately
the experimental neutrino data.
% for mass square differences and mixing angles.

%{\tt -- so see after the results --}\\
Another important issue is the renormalization group (RG) flow of
the neutrino parameters from the high energy scale where the
neutrino mass matrices are formed, down to their low energy
measured values.
%Indeed, in most of the phenomenological explorations, the
%relevant mass entries are determined in terms of the
One can attempt to determine the neutrino mass matrices from
experimental data directly at the weak scale. However, the Yukawa
couplings and other relevant parameters are not known at the
unification scale. A knowledge of
these quantities at the unification mass could provide a clue for
the structure of the unified or partially unified theory and the
exact (family) symmetries determining the neutrino mass matrices
at the GUT scale. Attempts to determine the neutrino mass
parameters in a top-bottom or bottom-up approach have recently
discussed in the
literature~\cite{Antusch:2005gp,Mei:2005qp,Ellis:2005dr}.

 In this paper we investigate further the neutrino mass spectrum
 of a model with gauge symmetry
$SU(4)\times SU(2)\times SU(2)\times U(1)_X$
  based on the 4-4-2 models~\cite{Antoniadis:1988cm,Allanach:1995ch,Allanach:1996hz},
  whose implications for quark and lepton masses were recently
investigated in~\cite{Dent:2004dn}. These models present several
attractive features.
 Firstly, only "small" Higgs representations are needed and these
 commonly arise in string models. Secondly, third generation fermion Yukawa couplings
 are unified \cite{AllanachKQuad} up to small corrections. Unification of gauge couplings
 is allowed and, if one assumes the model embedded in supersymmetric string, might be
 predicted \cite{LTracas}. Furthermore, the doublet-triplet splitting problem is
 absent.

 In string derived models a large number of neutral singlet fields carrying
 quantum numbers only under U$(1)_X$ appear in the spectrum of the effective field
 theory model. D- and F-flatness conditions require some of them to obtain
 non-zero vevs of the order of the U$(1)_X$ breaking scale. In the present model,
 in order to describe accurately the low energy neutrino
 data we introduce two such singlets charged under
 U$(1)_X$~\cite{Irges:1998ax}. The previous model \cite{Dent:2004dn} with one
 such singlet could easily give rise to a spectrum of light neutrinos with
 normal hierarchy and bi-large mixing. However, after study of the
 renormalization group (RG) evolution and unification it was found that the
 scale of light neutrino masses too large to be compatible with observation.
 If we impose the correct scale of light neutrino masses, then some heavy
 right-handed neutrinos (RHN) would have masses above the unification scale,
 which is incompatible with our effective field theory approach.

 Thus three {\em a priori}\/
 independent expansion parameters arise from the superpotential, two coming
 from the singlets and one from the Higgses which receive v.e.v.'s at the
 SU$(4)\times$SU$(2)_R$ breaking scale. In general, a nonrenormalizable
 operator may contain several products of the
 SU$(4)\times$SU$(2)_R$ breaking Higgses, and thus different contractions
 of gauge group indices are possible leading to different contributions
 depending on the Clebsch factors. We use a minimal set of nontrivial Clebsch
 factors to construct the Dirac mass matrices. Right-handed (SU$(2)_L$ singlet)
 neutrinos acquire Majorana masses through nonrenormalizable couplings to the
 U$(1)_X$-charged singlets and to Higgses, while light neutrinos will obtain
 masses via the see-saw mechanism.

 The renormalization group equations (RGEs)
 for the neutrino masses and mixing angles above, between and below the see-saw
 scales are solved numerically, for several sets of order 1 parameters which
 specify the heavy RHN matrix. In each case the results at low energy are
 consistent with current experimental data, and provide further predictions for
 the 1-3 neutrino mixing angle and for neutrinoless double beta decay.

% and the predictions of the model at low energy
% turns out to be in agreement with experiment.

\section{Description of the model}

In this section we present salient features of the
string inspired Pati-Salam model extended by a U$(1)_X$
family-symmetry, the total gauge group being
SU$(4)\times$SU$(2)_L\times$SU$(2)_R\times$U$(1)_X$.
The field content includes three copies of
$(4,2,1)+(\bar 4,1,2)$ representations to accommodate the three
fermion generations $F_i+\bar F_i$ ($i=1,2,3$),
\beq
F_i = \left( \begin{array}{cccc} u_i& u_i&u_i&\nu_i
 \\
 d_i& d_i&d_i&e_i
\end{array} \right)_{\alpha_i},\
\bar F_i = \left( \begin{array}{cccc} u_i^c&
u_i^c&u_i^c&\nu_i^c
 \\
 d_i^c& d_i^c&d_i&e_i^c
\end{array} \right)_{\bar\alpha_i} \label{fereps}\nonumber
\eeq
% $F_i$ includes all left-handed SM fermions (quarks and leptons),
% whilst $\bar F_i$ contains their right-handed partners, including the
% right-handed neutrinos.
where the subscripts $\alpha_i$, $\bar\alpha_i$ indicate the U$(1)_X$
charge.
%of the fermion generations
In order to break the Pati-Salam symmetry down to SM gauge group,
Higgs fields $H=(4,1,2)$ and $\bar H=(\bar 4,1,2)$ are introduced
\bea
H = \left( \begin{array}{cccc} u_H& u_H&u_H&\nu_H
 \\
 d_H& d_H&d_H&e_H
\end{array} \right)_{x},
\; \bar H = \left( \begin{array}{cccc} u_{\bar H}^c& u_{\bar
H}^c&u_{\bar H}^c&\nu_{\bar H}^c

 \\
 d_{\bar H}^c& d_{\bar H}^c&d_{\bar H}&e_{\bar H}^c
\end{array} \right)_{\bar x}\label{Hireps}\nonumber
\eea
which acquire vevs of the order $M_G$ along their neutral
components
\bea \vev{H}= \vev{\nu_H} = M_{G},\
\vev{\bar{H}}=\vev{\nu_{\bar{H}}} = M_{G}.
\eea
The Higgs sector also includes the $h=(1,\bar{2},2)$ field
which after the breaking of the PS symmetry is decomposed to
the two Higgs superfields of the MSSM.
Further, two $D=(6,1,1)$ scalar fields are introduced to give mass
to color triplet components of $H$ and $\bar{H}$ via the terms
$HHD$ and $\bar{H}\bar{H}D$~\cite{Antoniadis:1988cm}.

Finally, we
introduce two scalar singlet fields $\phi$, $\chi$, charged under
U$(1)_X$ whose vevs will play a crucial role in the fermion
mass matrices through non-renormalizable terms of the
superpotential. In the stable SUSY vacuum the two singlets obtain
vevs to satisfy the D-flatness condition including the anomalous
Fayet-Iliopoulos term~\cite{DineSW}. The anomalous D-flatness
conditions allow solutions where the vevs of the conjugate fields
$\bar\phi$ and $\bar\chi$ are zero and we will restrict our
analysis to this case. Note that in general a string model may
have more than two singlets and more than one set of Higgses $H_i$,
$\bar{H}_i$, with different U$(1)_X$ charge. All such fields may
in principle also obtain vevs,
%in accordance with the flatness conditions,
however we find that two of them are sufficient to give a set of
mass matrices in accordance with all experimental data. Hence we
consider any additional singlet vevs to be significantly smaller.
%than those of $\phi$, $\chi$.

The Higgses $H_i$, $\bar H_i$ may obtain masses through
$H\bar{H}\phi$, $H\bar{H}\chi$ and $H\bar{H}\phi\chi$ couplings.
However, in order to break the Pati-Salam group while preserving
SUSY we require that one $H$-$\bar{H}$ pair be massless at this
level. This ``symmetry-breaking'' Higgs pair could be a linear
combination of fields with different U$(1)_X$ charges, which
would in general complicate the expressions for fermion masses.
%(as will the presence of many $\phi_i$ fields).
The chiral spectrum is
%and its quantum numbers under the gauge symmetry of the model
summarized
%for convenience
in Table~\ref{spec}. We choose the charge of the Higgs field
$h$ to be $-\alpha_3-\bar\alpha_3$ so that that the 3\rd
generation coupling $F_3\bar{F}_3h$ is allowed at tree-level.

\begin{table}[here]
 \caption{Field content and U$(1)_X$ charges}\label{table1}
\begin{centering}
\begin{ruledtabular}
\begin{tabular}{c|cccc}
         &  SU$(4)$ & SU$(2)_L$ & SU$(2)_R$ & U$(1)_X$  \\
         \hline
$F_i$      &     4    &     2     &     1     &  $\al_i $ \\
$\bar{F}_i$& $\bar{4}$&     1     & $\bar{2}$ & $\bar{\al}_i$ \\
$H$        &     4    &     1     &     2     &   $x$     \\
$\bar{H}$  & $\bar{4}$&     1     & $\bar{2}$ & $\bar{x}$ \\
$\phi$     &     1    &     1     &     1     &   $z$     \\
$\chi$     &     1    &     1     &     1
  &   $z'$     \\
$h$        &     1    & $\bar{2}$ &     2     & $-\al_3-\bar{\al}_3$ \\
$D_1$      &     6    &     1     &     1     &  $-2x$    \\
$D_2$      &     6    &     1     &     1     & $-2\bar{x}$
\end{tabular}
\end{ruledtabular}
\end{centering}\label{spec}
\end{table}

We now turn to the terms in the superpotential which can give rise
to fermion masses.
%Charged fermions obtain only Dirac type mass terms, whilst neutral
%ones may obtain Dirac and Majorana masses.
Dirac type mass terms arise after electroweak symmetry-breaking from
couplings of the form
\begin{multline}
 {\cal W}_{D} = y_0^{33} F_3\bar{F}_3h + F_i\bar{F}_j h
 \left( \sum_{m>0} y_m^{ij} \left(\frac{\phi}{M_U}\right)^m+ \right. \\
\left.\sum_{m'>0} y_{m'}^{ij} \left(\frac{\chi}{M_U}\right)^{m'}+
\sum_{n>0} {y'}_n^{ij}  \left(\frac{H\bar{H}}{M_U^2} \right)^n+   \right.\\
\left. \sum_{k,\ell>0} y_{k,\ell}^{ij}
\left(\frac{\phi}{M_U}\right)^k
\left(\frac{\chi}{M_U}\right)^{\ell} +  \sum_{p,q>0} y_{p,q}^{ij}
\left(\frac{H\bar{H}}{M_U^2} \right)^p
\left(\frac{\phi}{M_U}\right)^{q}   \right.\\
 \left.+  \sum_{r,s>0} y_{r,s}^{ij}
\left(\frac{H\bar{H}}{M_U^2}
 \right)^r
\left(\frac{\chi}{M_U}\right)^{s}
  + \cdots \right) \label{Dsup}
\end{multline}
%where $m$, $m'$, $n,k,l,p,q,r,s$ are appropriate integers.
Apart from the heaviest generation, all masses arise at
non-renormalizable level, suppressed by powers of the
fundamental scale or unification scale $M_U$. The couplings
$y_{m,m'}^{ij}$, ${y'}_n^{ij}$ {\em etc.}\ are non-vanishing
and generically of order $1$ whenever the U$(1)_X$ charge of
the corresponding operator vanishes, thus:
 \begin{multline}
 \alpha_i-\alpha_3 +\bar{\alpha}_j-\bar{\alpha}_3 =  \{ -m z, -m' z' , -n (x+\bar{x}), \\
 -k z - \ell z' , -p (x+\bar{x})  -q z , -r (x+\bar{x}) -s z' \}. \nonumber
 \end{multline}
Other higher-dimension operators may arise by multiplying any term
by factors such as $(H\bar{H})^\ell\phi^s/M_U^{2\ell+s}$
%for positive integers $\ell$ and $s$ such that
where $\ell(x+\bar{x})+s\,z=0$.
Such terms are negligible unless the leading term vanishes.

Neutrinos
%al fermions (neutrinos)
may in addition receive also
Majorana type masses. These arise from the operators
\begin{multline}
 W_{M} = \frac{\bar{F}_i\bar{F}_j HH}{M_U} \left( \mu_0^{ij} + \sum_{t>0}
\mu_t^{ij} \left(\frac{\phi}{M_U}\right)^t  \right. \\
\left.+ \sum_{t'>0} \mu_{t'}^{ij}
\left(\frac{\chi}{M_U}\right)^{t'} + \sum_{w>0} {\mu'}_w^{ij}
\left(\frac{H\bar{H}}{M_U^2}\right)^w \right. \\
\left.+ \sum_{k',\ell'>0} \mu_{k',\ell'}^{ij}
\left(\frac{\phi}{M_U}\right)^{k'}
\left(\frac{\chi}{M_U}\right)^{\ell'}  \right. \\
\left.+
 \sum_{p',q'>0} \mu_{p',q'}^{ij}  \left(\frac{H\bar{H}}{M_U^2}
 \right)^{p'}
\left(\frac{\phi}{M_U}\right)^{q'} \right. \\
\left.+  \sum_{r,s>0} \mu_{r',s'}^{ij}
\left(\frac{H\bar{H}}{M_U^2}
 \right)^{r'}
\left(\frac{\chi}{M_U}\right)^{s'}
  + \cdots
\right).
 \label{MM}
\end{multline}
% where $t$, $t'$ , $w,k',\ell'p',r',s'$ are appropriate integers.
 Couplings of this type are non-vanishing whenever the following
 conditions are satisfied:
\begin{multline}
 \bar{\alpha}_i+\bar{\alpha}_j +2x = \{ - tz, -t' z', -w (x+\bar{x}), \\
-k' z - \ell' z' -p' (x+\bar{x}) -q' z, -r' (x+\bar{x}) -s' z'
\}. \nonumber
\end{multline}

\section{Fermion mass matrices}

\subsection{General structure}
As can be seen from the superpotential Yukawa couplings
(\ref{Dsup}) and (\ref{MM}), three different expansion parameters
appear in the construction of the fermion mass matrices.
These are
\bea
\eps\equiv \frac{\vev{\phi}}{M_U},\ \ \
\eps'\equiv \frac{M_{\rm G}^2}{M_U^2},\ \ \
\eps''\equiv \frac{\vev{\chi}}{M_U} \label{expanspar}
\eea
given $\vev{H\bar H}= M _{G}^2$.
Note that, for non-renormalizable Dirac mass terms involving
several products of $H\bar{H}/M_U^2$, the gauge group indices may
be contracted in different ways \cite{Allanach:1996hz}.  This can
lead to different contributions to the up, down quarks and charged
leptons, depending on the Clebsch factors $C^{ij}_{n(u,d,e,\nu)}$
multiplying the effective Yukawa couplings.  Also, although the
Clebsch coefficient for a particular operator $O_n$ may vanish at
order $n$, the coefficient for the operator $O_{(n+p);q}$
containing $p$ additional factors $(H\bar{H})$ and $q$ factors of
$\phi$ and/or $\chi$ is generically nonzero.

In our analysis we wish to estimate the effects of the second
singlet ($\chi$) contributions on the neutrino sector as compared
to the analysis presented in~\cite{Dent:2004dn} without affecting
essentially the results in the quark sector.
In order to obtain a set of fermion mass matrices with the minimum
number of new operators, we assume fractional $U(1)_X$ charges
for $H,\bar H$ and $\chi$ fields, while the combination $H\bar H$
and the singlet $\phi$ are assumed to have integer charges.
% under $U(1)_X$.
Thus $\alpha_i$, $\bar{\alpha}_i$, $x+\bar x$ and $z$ are
integers, while $z'$, $x$ and $\bar{x}$ are fractional. As a
result, the Dirac mass terms involving vevs of $\chi$ are
expected to be subleading compared to other terms. Suppressing
higher-order terms involving products of $\eps,\eps'$
and $\eps''$, the Dirac mass terms at the unification scale are
\begin{align}
m_{ij} &\approx &\delta_{i3}\delta_{j3}m_3  + \left( \eps^{m} +
(\eps'')^{m'}+ C_{ij} (\eps')^{n} \right)\,v_{u,d}
\end{align}
where $m_3\equiv v_{u,d} y_0^{33}$, with $v_u$ and $v_d$ being the
up-type and down-type Higgs vevs respectively, and we omit the
order-one Yukawa coefficients $y^{ij}_m$ {\em etc.}\/ for simplicity.

%For convenience, defining $\hat{m}= -( \alpha_i-\alpha_3
%+\bar{\alpha}_j-\bar{\alpha}_3)$,  $\hat{m'}= -(
% \alpha_i
%-\alpha_3
%+\bar{\alpha}_j-\bar{\alpha}_3)$ and $\hat{n}= -(\alpha_i-\alpha_3
%+\bar{\alpha}_j-\bar{\alpha}_3)$ we can rewrite the constraints as
%$\hat{m} =mz $, $\hat{m'} =m'z' $ and $\hat{n}= n(x+\bar{x})$. The
%signs of $\hat{m}$, $\hat{m'}$ and $\hat{n}$ must be the same as
%$z$,$z'$ and $x+\bar{x}$ respectively for the mass term to exist.
%So (\ref{expanspar}) can be
% written
% \bea
%  \epsilon^{|\hat{m}|}\equiv \left(\frac{ \langle{\phi}\rangle}{M_U}\right)^{m}
%  , \ \
%  \epsilon'^{|\hat{n}|}\equiv \left(\frac{M_{\rm
%  G}^2}{M_U^2}\right)^{n}, \ \
%  \epsilon''^{|\hat{m'}|}\equiv \left(\frac{ \langle{\chi}\rangle}{M_U}\right)^{m'}
%  \eea

 The Majorana mass terms are proportional to the combination $HH$
 (see Eq.~(\ref{MM})) which has fractional $U(1)_X$ charge. Thus,
 terms proportional to $\chi/M_U$ become now important for the structure
 of the mass matrix. The general form of the Majorana mass matrix is then
  \begin{multline}
 M_N \approx M_R \left( \mu_t^{ij}\epsilon^{t} +
 \mu_{t'}^{ij}(\epsilon'')^{t'} + {\mu'}_{w}^{ij} (\epsilon')^{w}  \right. \nonumber   \\
\left.  + \mu_{k',l'}^{ij} \epsilon^{k'} (\epsilon'')^{l'} +
\mu_{p',q'}^{ij} (\epsilon')^{p'} \epsilon^{q'} + \mu_{r',s'}^{ij}
(\epsilon')^{r'} (\epsilon'')^{s'}
 \right)
 \end{multline}
where we define $M_R\equiv M_{\rm G}^2/M_U\equiv\eps'M_{U}$.

%Analogously,
% we defined $\hat{p}= -(\bar{\alpha}_i +\bar{\alpha}_j +2x)$,
% $\hat{q}= -(\bar{\alpha}_i +\bar{\alpha}_j +2x)$ and
% $\hat{r}= -(\bar{\alpha}_i +\bar{\alpha}_j +2x)$.

\subsection{Specific choice of $U(1)_X$ charges}
Before we proceed to a specific, viable set of mass matrices,
we first make use of the observation~\cite{Dent:2004dn}
that the form of the fermion mass terms above is invariant
under the shifts
 \beq
\alpha_i \rightarrow \alpha_i + \zeta, \
\bar{\alpha}_i \rightarrow \bar{\alpha}_i + \bar{\zeta}, \
x \rightarrow x - \bar{\zeta}, \
\bar{x} \rightarrow \bar{x} + \bar{\zeta} \label{zeta}
 \eeq
so that we are free to assign $\alpha_3=\bar\alpha_3=0$. We
further fix $x+\bar{x}=1$ and $z=-1$; we will choose the
values of $x$ and $z'$ to be fractional such that the v.e.v.\
of $\chi$ only affects the overall scale of neutrino masses,
as explained below.
%and $z'=-\frac 35$. The
%resulting $U(1)_X$ charges are presented in Table \ref{charges}.
%\begin{table}
%\begin{center}
%\begin{tabular}{|c|ccccccccccc|}
%    \hline
%  field & $F_1$ & $F_2$ & $F_3$ & $\bar{F}_1$ & $\bar{F}_2$ &
%$\bar{F}_3$ &
%$h$ & $H$ & $\bar H$ & $\phi$ & $\chi$  \\
%    \hline
%U$(1)_X$& -4 &  -3 &  0 &
% -2 &  1 & 0 &
% 0 & $x$ &  $\bar{x}$ & -1 & $ -3/5$ \\
%    \hline
%\end{tabular}
%\end{center}
%\caption{\label{charges} Specific choice of U(1)$_\mathrm{X}$ charges}
%\end{table}

The charge entries of the common Dirac mass matrix
for quarks, charged leptons and neutrinos
%, and the neutrino Majorana mass matrix,
are then
\beq
\label{chargem} Q_X[M_D] =
%\left( \begin{array}{ccc}
\begin{pmatrix}
-6 & -3 & -4 \\ -5 & -2 & -3 \\ -2 & 1 & 0
%\end{array} \right)
\end{pmatrix},
\eeq
and the charge matrix for heavy neutrino Majorana masses is
\beq
Q_X[M_N] = 2x +
%\left( \begin{array}{ccc}
\begin{pmatrix}
-4 & -1 & -2 \\ -1 & 2 & 1 \\ -2 & 1 & 0
%\end{array} \right)
\end{pmatrix}. \label{pmatrix}
\eeq
Now, we relate $\eps,\eps',\eps''$ with a single expansion
parameter $\eta$, assuming the relations
 \beq
\eps =b_1 \sqrt{\eta},\ \eps''\equiv b_2 \eta,\ \eps'=\sqrt{\eta}
\label{rel}
 \eeq
where $b_1$, $b_2$ are numerical coefficients of order one. Then
the effective Yukawa couplings for quarks and leptons may include
terms
\beq Y_f^{ij}=
%y_f^{ij}
 b_1^m \eta^{m/2} + b_1^{m+1} \eta^{1+m/2}
+ C_f^{ij}\eta^{n/2} + b_1 \eta^{1+n/2} + \cdots
 \label{yukawas}
\eeq
with $f=u,d,e,\nu$, up to order 1 coefficients $y_f^{ij}$.
Which of these terms survives, depends on
the sign of the charge of the corresponding operator. For a
negative charge entry, the first two terms are not
allowed and only the third and fourth contribute. Further,
if a particular $C_{f}^{ij}$ coefficient is zero, then we
consider only the fourth term.

Therefore, we need to specify the Clebsch-Gordan coefficients
$C^{ij}_f$ for the terms involving powers of
%the expansion parameter
$\langle{H\bar H}\rangle/M_U^2$.
These coefficients
%$C_f^{ij}$
could be found if the fundamental theory
%with all the symmetries
was completely specified at the unification or string scale. In
the absence of a specific string model, here we present a minimal
number of operators which lead to a simple and viable set of mass
matrices. Up to possible complex phases, we choose
$C_d^{12}=C_d^{22}=\frac 13$, $C_u^{23}=3$ and
$C_u^{11}=C_u^{12}=C_u^{21}=C_u^{22}=C_u^{31}=C_{\nu}^{22}=C_{\nu}^{31}=0$
with all others being equal to unity. The effective Yukawa
matrices at the GUT scale obtained under the above assumptions are
\begin{gather}
 Y_u = \begin{pmatrix}
 b_1 \eta^{4} & b_1 \eta^{5/2} & \eta^2 \\
  b_1 \eta^{7/2} & b_1 \eta^{2} & 3\eta^{3/2} \\
 b_1 \eta^{2} & b_1 \eta^{1/2} & 1
\end{pmatrix},\
 Y_d = \begin{pmatrix}
 \eta^3 & \frac{\eta^{3/2}}{3} & \eta^2 \\
 \eta^{5/2} & \frac{\eta}{3} & \eta^{3/2} \\
 \eta & b_1  \eta^{1/2} & 1
\end{pmatrix}, \nonumber  \\
 Y_e = \begin{pmatrix}
 \eta^3 & \eta^{3/2} & \eta^2 \\
 \eta^{5/2} & \eta & \eta^{3/2}  \\
 \eta & b_1  \eta^{1/2} & 1
\end{pmatrix},\
 Y_{\nu} = \begin{pmatrix}
 \eta^3 & \eta^{3/2} & \eta^2 \\
 \eta^{5/2} & b_1 \eta^{2} & \eta^{3/2}  \\
 b_1 \eta^{2} & b_1 \eta^{1/2} & 1
\end{pmatrix}
\end{gather}
where we suppress order one coefficients.  The quark sector as
well as the neutrino sector were studied
 in~\cite{Dent:2004dn}. However, full renormalization group
 effects
%above between and below the seesaw scales
 were not calculated for the neutrino sector and as it turns out
 one singlet is inadequate to accommodate the low energy data.
 Consequently, we introduced the second singlet $\chi$, with
 fractional charge, whose v.e.v.\ affects only the overall scale
 of neutrino masses.

The desired matrix for the right handed
 Majorana neutrinos may result from more than one choice of
 charge for the $H$ field and the $\chi$ singlet field. These can
 be seen in Table \ref{posscharges}.
\begin{table}
\begin{center}
\begin{tabular}{|c|c|c|c|c|c|c|c|}
    \hline
  $Q_X[H]$ & $-\frac 23$ & $-\frac 43$ & $-\frac 53$ & $-\frac 65$ & $-\frac 78$ &
$-\frac 14$ &
$-\frac 34$   \\
    \hline
$Q_X[\chi]$& $-\frac 53$ &  $-\frac 13$ & $\frac 13$ &
 $-\frac 35$ &  $-\frac 54$ & $-\frac 52$ &
 $-\frac 32$  \\
    \hline
\end{tabular}
\end{center}
\caption{\label{posscharges} Possible choices for the
U(1)$_\mathrm{X}$ charges of the $H$ and $\chi$ fields.}
\end{table}
 We choose the $H$ charge to be $x=-\frac 65$ so that $2\,x$
 is non-integer, and set the
%$U(1)_X$ charge of the
 $\chi$ singlet charge to $-\frac 35$. The analysis for the
 quarks and charged leptons remains the as in~\cite{Dent:2004dn}
 since operators with nonzero powers $\chi^r$
%of the form $F_i\bar{F}_j h(\frac{H\bar{H}}{M_U^2})^n (\frac{\chi}{M_U})^r$
 do not exist for powers $r<5$ and are
 negligible compared to the leading terms. %$F_i\bar{F}_j h(\frac{H\bar{H}}{M_U^2})^n$.

 With these assignments, the charge entries of the
 heavy Majorana matrix Eq.~(\ref{pmatrix}) are:
 \beq
 Q_X[M_N] =\left( \begin{array}{ccc}
 -\frac{32}{5}  & -\frac{17}{5} & -\frac{22}{5} \\
  -\frac{17}{5} & -\frac{2}{5}& -\frac{7}{5} \\
   -\frac{22}{5} & -\frac{7}{5}  & -\frac{12}{5}
\end{array} \right).
 \eeq
% For the neutrino sector we make the following observations.
 Due to the fractional $U(1)_X$ charges
%of the operators responsible for generating neutrino masses,
 contributions from $\phi$ or $H\bar{H}$
 alone vanish. However, we also have the singlets $H\bar{H}\chi/M_U^3$ with
 vev $b_2\, \eta^{3/2}$ and $\phi \chi /M_U^2$ with a vev $b_1 b_2 \,\eta^{2}$,
 while for some entries one may have to consider higher order
 terms since the leading order will be vanishing.
 In Table \ref{operators} we explicitly write the operator for
every entry of $M_N$.
  The Majorana right-handed neutrino
 mass matrix is then
 \beq \label{MRHN}
 M_N=\left( \begin{array}{ccc}
 \mu_{11} \eta^{9/2}  & \mu_{12}\eta^{3} & \mu_{13}\eta^{7/2}\\
 \mu_{12}\eta^{3} & \mu_{22}\eta^{3/2}& \mu_{23}
 \eta^{2} \\
  \mu_{13} \eta^{7/2} & \mu_{23}\eta^{2}  &  \eta^{5/2}
\end{array} \right) b_2 M_R
\eeq
 with $M_R = \eps' M_U = \sqrt{\eta} M_U $.

Having defined the Dirac and heavy Majorana mass matrices for the
neutrinos, it is straightforward to obtain the light Majorana mass
matrix
%, relevant to neutrino experimental data,
from the see-saw formula
\beq \label{seesaw}
 m_\nu = - m_{D\nu} M_N^{-1} m_{D\nu}^T
\eeq
at the GUT scale.

\begin{table}
\begin{center}
\begin{tabular}{|c|c|c|}\hline
 $M_N$ entry & Operator & vev\\
\hline $  M_N^{11}$ & $ \left(\frac{H\bar{H}}{M_U^2}\right)^6
\frac{\chi}{M_U} $
 & $b_2 \eta^{9/2}$\\
\hline $  M_N^{12}$ & $  \left(\frac{H\bar{H}}{M_U^2}\right)^4
\frac{\chi}{M_U} $
 & $b_2 \eta^{3}$\\
\hline $  M_N^{13}$ & $ \left(\frac{H\bar{H}}{M_U^2}\right)^5
\frac{\chi}{M_U} $
 & $b_2 \eta^{7/2}$ \\
\hline $  M_N^{22}$ & $ \left(\frac{H\bar{H}}{M_U^2}\right)
\frac{\chi}{M_U} $
 & $b_2  \eta^{3/2}$\\
\hline $  M_N^{23}$ & $ \left(\frac{H\bar{H}}{M_U^2}\right)^2
\frac{\chi}{M_U} $
 & $b_2 \eta^{2} $ \\
\hline $  M_N^{33}$ & $ \left(\frac{H\bar{H}}{M_U^2}\right)^3
\frac{\chi}{M_U} $
 & $b_2 \eta^{5/2}$\\
\hline
\end{tabular}
\end{center}
\caption{\label{operators}Operators producing the Majorana right
handed neutrino matrix $M_N$.}
\end{table}
%\left(\frac{H\bar{H}}{M_U^2}\right)^2 \left(\frac{\phi}{M_U}
%\right) \left(\frac{\chi}{M_U} \right)^3 }
%$b^3(\sqrt{\eta})^2 \eta \eta^3 =b^3\eta^{5}$

\subsection{Setting the expansion parameters}

Given the fermion mass textures in terms of the $U(1)_X$
charges and expansion parameters, we need now to determine the
values of the latter in order to obtain consistency with the low
energy experimentally known quantities (masses and mixing angles).
%Note that in our simplified approach, all three expansion
%parameters are related through .
Note that the coefficient $b_2$ defined in Eq.~(\ref{rel}) will
determine the overall neutrino mass scale through Eq.~(\ref{MRHN}).
%, while the necessity of $b_1$ stems from the desire of appropriate
%coefficients in the dirac and majorana matrices.
%, the
%relation $\eps\approx b^{-1} \,\eps''\equiv \eta$ implies that the
%magnitude of the $\phi$ and $\chi$ vevs are related by the
%coefficient $b$ which, will be determined by the neutrino mass
%scale.

Consistency with the measured values of quark masses and mixings fixes the value
of $\eta\approx 5\times 10^{-2}$: for example the CKM mixing angle $\theta_{12}$
is given by $\sqrt{\eta}\approx 0.22$ up to small corrections~\cite{Dent:2004dn}.
Hence the ratio of the SU$(4)$ breaking scale $M_G$ to the fundamental scale
$M_U$ is also fixed through $\frac{M_G^2}{M_U^2} =\sqrt{\eta} \approx 0.22$: the
Pati-Salam group is unbroken over only a small range of energy. We perform a
renormalization group analysis in order to check the consistency of this
prediction with the low-energy values of the gauge couplings $\alpha_s$,
$\alpha_{em}$ and the weak mixing angle $\sin^2\theta_W$ \cite{Eidelman}
 \beq
 \sin^2\theta_W =0.23120,\ \alpha_3 = 0.118\pm 0.003,\ a_{em} = \frac{1}{127.906}.
 \nonumber
 \eeq
 If the underlying model at $M_U$ has a single unified gauge
 coupling, then $M_G$ is fixed to be just below the unification
 scale according to the analysis of gauge coupling unification
 in the MSSM. Because of this fact, the low energy measured range
 for $\alpha_3$ affects the unification of the gauge couplings.
 Thus, we add the following extra states
 \beq
 h_L = (1,2,1),\ \  h_R=(1,1,2)
 \nonumber
 \eeq
 which are usually present in a string spectrum~\cite{Antoniadis:1988cm}.
 It turns out that we need 4 of each of these extra states for $M_G = 9.32
 \times 10^{15}$ GeV to be consistent with the value of $\eps'$ deduced
 from quark mass matrices.

\begin{figure}[tbh]
\centering
\includegraphics[scale=.88]{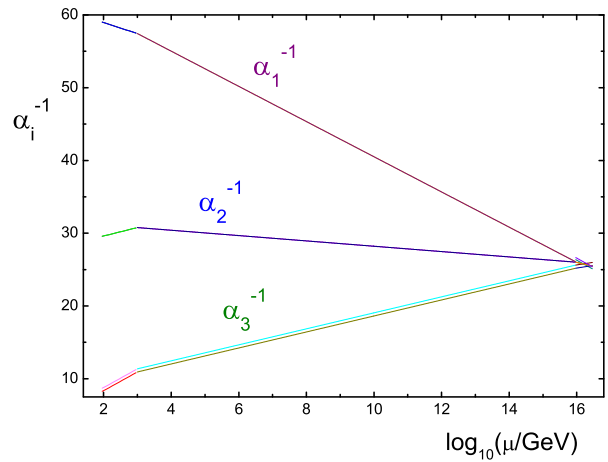}
\caption{Evolution of the gauge couplings. The two lines for $\alpha_3$
indicate the range of initial conditions at $M_Z$.}
 \label{ggevo}
\end{figure}

\begin{figure}[tbh]
\centering
\includegraphics[scale=.88]{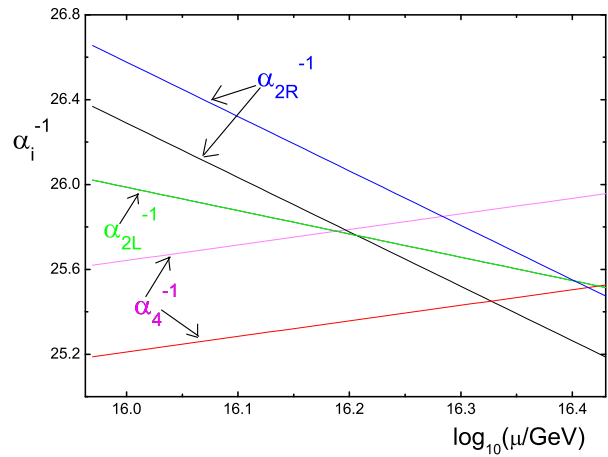}
\caption{\label{closeup} Close-up of the gauge couplings in the
  Pati-Salam energy region.} %for the $\alpha_3 (M_Z)$ range. }
\end{figure}

\begin{figure}[tbh]
\centering
\includegraphics[scale=.88]{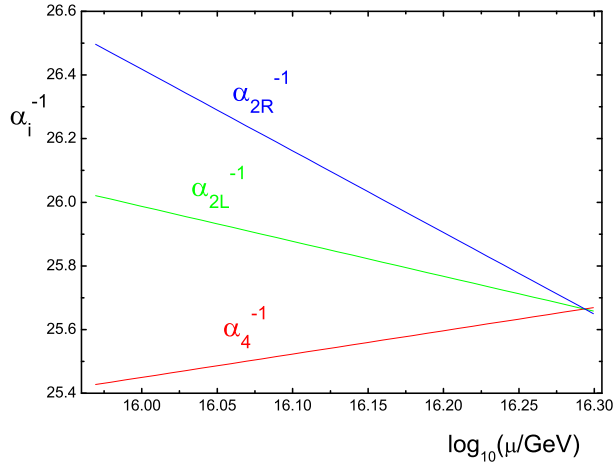}
\caption{\label{psuni} The unification point of Pati-Salam gauge
couplings for $\alpha_3 (M_Z)= 0.1176$. }
\end{figure}

 In Figure \ref{ggevo}
 we plot the evolution of the gauge couplings from $M_Z$ to $M_U$.
  In Figure \ref{closeup}
 we show in more detail the evolution of the gauge couplings in the
 Pati-Salam energy region. The two bands for the $\alpha_4$
 and $\alpha_{2R}$ couplings are due to strong coupling
 uncertainty  at $M_Z$. For $\alpha_3 (M_Z)=0.1176$, as can
 be seen from Figure~\ref{psuni}, we obtain $M_U=1.96 \times 10^{16}$ GeV.

%%%%%%%%%%%%%%%%%%%%%%%%%%%Figures%%%%%%%%%%%%%%%%%%%%%%%%%%%%%%%%%%%%%%%
%%%%%%%%%%%%%%%%%%%%%%%%%%%Figures%%%%%%%%%%%%%%%%%%%%%%%%%%%%%%%%%%%%%%%

The remaining parameters to be determined are $b_1$, $b_2$ and
$\mu_{ij}$ of the right handed neutrino mass matrix.  We find that
$M_N$ is proportional to $b_2$, $M_N\approx b_2 M_R M_N' $, thus
$b_2$ is related to the scale of the light matrix $m_{\nu}$. Also,
 the choice $b_1 \approx 1.1$ leads to agreement with the data, while
implying $\eps = 1.1 \sqrt{\eta} \approx 0.25$.

\section{Running of neutrino masses and mixing angles}

One of the problems one encounters when searching for a specific
 mass matrix for the light neutrinos via the see-saw formula is
 the effects induced by the renormalization group equations. The
 low energy neutrino data
%stemming from a particular set up
 could be considerably different from the results at the see-saw scale.
 The
%differential equations for the
 running of neutrino
 masses and mixing angles has been extensively discussed for
 energies below the seesaw scales~\cite{below,Antusch:2002rr,Antusch:2003kp}
 as well as above~\cite{above,Antusch:2005gp,Mei:2005qp}. The
 running of the effective neutrino mass matrix $m_{\nu}$ above and
 between the see-saw scales is split into two terms,
 \beq
    m_\nu \,=\,
    -\frac{v^2}{4} \left(
    \, \accentset{(n)}{\kappa} +
    2 \, \accentset{(n)}{Y}_\nu \accentset{(n)}{M}^{-1}
     \accentset{(n)}{Y}_\nu^T \right) \;.
 \label{mnr}
 \eeq
where $\kappa$ is related to the coefficient of the effective 5
dimensional operator $LLh_uh_u$, $(n)$ labels the effective field
theories with $M_n$ right handed neutrino integrated out ($M_n \ge
M_{n-1} \ge M_{n-2}, \ldots$)
 and $ \accentset{(n)}{Y}_\nu$ are the neutrino couplings
 at an energy scale $M$ between two RH neutrino masses $M_n \ge M
 \ge M_{n-1}$, while $\accentset{(n)}{Y}_\nu=0$ below the
 lightest RH neutrino mass. These effective parameters
% in (\ref{mnr})
 govern the evolution below the highest seesaw scale
 and obey the differential
 equation~\cite{below,Antusch:2002rr,Antusch:2003kp}
\begin{multline}
\label{eq} 16 \pi^{2} \frac{d \effn{n} X}{dt} = ( Y_e Y_e^\dagger
+ \effn{n} Y_\nu \effn{n} Y_\nu^\dagger)^T  \effn{n} X + \effn{n}
X ( Y_e Y_e^\dagger  + \effn{n} Y_\nu \effn{n} Y_\nu^\dagger)^T
 \\
 + (2 \mrm{Tr} (\effn{n} Y_\nu \effn{n} Y_\nu^\dagger +
3 Y_u Y_u^\dagger) -6/5 g_1^2 -6 g_2^2) \effn{n} X
\end{multline}
 where $X=\kappa$, $Y_\nu M^{-1} Y_\nu^T$.
 The RGEs have been solved both numerically and also
 analytically~\cite{Antusch:2003kp,Antusch:2005gp,Mei:2005qp}.
 Numerically, below the lightest heavy RH neutrino mass large
 renormalization effects can be experienced only in the case of
 degenerate light neutrino masses for very large $\tan\beta$~\cite{Ellis:1999my,Antusch:2003kp}.
 Above this mass things are more complicated due to the
 non-trivial dependence of the heavy neutrino mass couplings,
 unless $M_{\nu^c}$ is diagonal. For the leptonic mixing
 angles, in the case of normal hierarchy relevant to our
 model, one expects negligible effects for the solar mixing
 angle while $\theta_{13}$ and $\theta_{23}$ are expected
 to run faster~\cite{Ellis:2005dr}.
% Therefore, the quark-lepton complementarity relation (QLC) is stable.

 On the other hand, studying the analytical expressions
 obtained after approximation, exactly the opposite behavior
 is predicted and the solar mixing angle receives larger
 renormalization effects than $\theta_{13}$ or $\theta_{23}$.
 However, possible cancellations may occur and enhancement or suppresion factors
 may appear, thus the numerical solutions may differ considerably
 from these estimates.

In our string-inspired model the Dirac and heavy Majorana mass
matrices at the unification scale are parametrized in terms of
order-1 superpotential coefficients $\mu_{ij}(M_U)$
%at the unification scale
whose exact numerical values are not known. The flavour
structure at the unification scale might also be different from
that at the electroweak scale $M_Z$. Thus, even if the Yukawa
parameters are determined at $M_Z$, to understand the structure
of the theory at $M_U$, and consequently any possible family
symmetry, we would certainly need the parameter values at $M_U$.

In this section we study the renormalization group flow of the
neutrino mass matrices ``top-down'' from the Pati-Salam scale
$M_G$ to the weak scale. We
choose sets of values of the undetermined order 1 coefficients
at the high scale and run the renormalization group equations
down to $M_Z$ where we calculate $\Delta m_{\nu_{ij}}^{2}$ and
$\theta_{\nu_{ij}}$ and compare them with the experimental values.
Study of a bottom-up approach has been
performed~\cite{Ellis:2005dr} and we will compare our results to
this work. The renormalization group analyses of the neutrino
parameters, successively integrating out the right handed
neutrinos, is performed using the software packages
REAP/MPT~\cite{Antusch:2005gp}.

$\bullet$ We generate appropriate numerical values for the
coefficients $\mu_{11}$, $\mu_{12}$, $\mu_{13}$, $\mu_{22}$,
$\mu_{23}$, so that after the evolution of $m_{\nu}$ to low energy
we obtain values in agreement with the experimental data. The
coefficient $\mu_33$ is set to unity (which can always be
done by adjusting the value of $b_2$).
Experimentally acceptable solutions can be seen in Table
\ref{nummuij}. In Table \ref{dmthetanum} we present the resulting
values of $\theta_{ij}$ and $\Delta m_{ij}^2$ at the scale $M_Z$.
The mass-squared differences lie in the ranges
 $\Delta m_{atm}^2=[1.33-3.39]\times 10^{-3}$eV$^2$, $\Delta
m_{sol}^2=[7.24-8.85]\times 10^{-5}$eV$^2$. These are
consistent with the experimental data ${\Delta m_{atm,exp}^2}=
[1.3-3.4]\times 10^{-3}$eV$^2$ and ${\Delta m_{sol,exp}^2}=
[7.1-8.9]\times 10^{-5}$eV$^2$. The mixing angles are also
found in the allowed ranges $\theta_{12}=[29.4-37.6]$,
$\theta_{23}=[36.9-51.0]$ and $\theta_{13}=[0.86-12.50]$.

\begin{table}
\begin{center}
\begin{tabular}{|c|c|c|c|c|c|}
\hline Solution & $\mu_{11}$& $\mu_{12}$ & $\mu_{13}$ & $\mu_{22}$ & $\mu_{23}$ \\
\hline
1.  & 0.10535 & 0.10972 & 0.86012 & 0.10491 & 0.91014 \\
\hline
2.  & 0.11939 & 0.10954 & 0.80912 & 0.10683 & 0.93832 \\
\hline
3.  & 0.10392 & 0.11787 & 0.97796 & 0.10512 & 0.98749 \\
\hline
4.  & 0.09143 & 0.10962 & 0.87616 & 0.10063 & 0.93798 \\
\hline
5.  & 0.12697 & 0.12745 & 0.99860 & 0.11652 & 0.99980 \\
\hline
6.  & 0.10920 & 0.09638 & 1.00975 & 0.10238 & 0.93561 \\
\hline
7.  & 0.10124 & 0.11682 & 0.98568 & 0.10688 & 0.99156 \\
\hline
8.  & 0.12358 & 0.09514 & 0.99580 & 0.10434 & 0.95646 \\
\hline
9.  & 0.13006 & 0.11973 & 1.02235 & 0.10378 & 0.89460 \\
\hline
10. & 0.12665 & 0.12137 & 1.00029 & 0.10695 & 0.91578 \\
\hline
\end{tabular}
\end{center}
\caption{Numerical values of parameters $\mu_{11}$,
$\mu_{12}$, $\mu_{13}$, $\mu_{22}$, $\mu_{23}$ at $M_G$.}
\label{nummuij}
\end{table}

\begin{table}
\begin{center}
\begin{tabular}{|c|c|c|c|c|c|}
\hline Soln. & $\Delta m_{atm}^2 \cdot 10^{3}$ & $\Delta m_{sol}^2 \cdot 10^{5}$ &
$\theta_{12}$ & $\theta_{13}$ & $\theta_{23}$ \\
\hline
1.  & 2.7149 & 7.9621 & 29.442 & 3.9859 & 44.114 \\
\hline
2.  & 2.3145 & 7.9514 & 34.289 & 12.507 & 51.047 \\
\hline
3.  & 1.8978 & 8.6141 & 30.560 & 0.8656 & 46.230 \\
\hline
4.  & 3.0062 & 8.3217 & 34.347 & 1.8512 & 44.333 \\
\hline
5.  & 3.3905 & 7.2468 & 30.245 & 2.9355 & 36.900 \\
\hline
6.  & 3.2459 & 7.5351 & 34.296 & 1.3701 & 46.947 \\
\hline
7.  & 2.0171 & 7.9464 & 34.432 & 1.0086 & 50.279 \\
\hline
8.  & 1.3321 & 7.9060 & 37.646 & 6.1490 & 43.067 \\
\hline
9.  & 2.4867 & 8.8561 & 29.592 & 5.6007 & 42.970 \\
\hline
10. & 2.1652 & 7.8869 & 29.189 & 3.1512 & 37.220 \\
\hline
\end{tabular}
\end{center}
 \caption{Values of the physical parameters $\Delta
m^2_{atm}$, $\Delta m^2_{sol}$, $\theta_{12}$, $\theta_{13}$,
$\theta_{23}$ at $M_Z$ (mass units eV$^2$).}
 \label{dmthetanum}
\end{table}

\begin{figure}[tbh]
\centering
\includegraphics[scale=0.68]{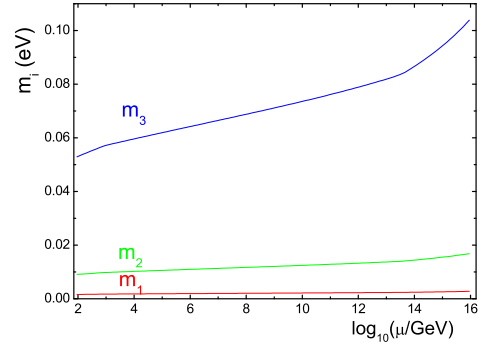}
\caption{The running of the light neutrino masses.}
 \label{lightneum}
\end{figure}

\begin{figure}[tbh]
\centering
\includegraphics[scale=0.68]{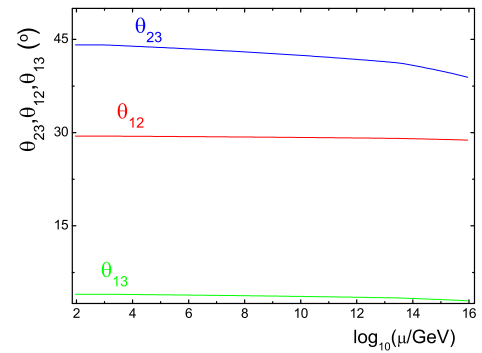}
\caption{The evolution of the mixing angles.}
 \label{anglesevolatex}
\end{figure}

$\bullet$ In figure (\ref{lightneum}) we plot the running of the three light
neutrino Majorana masses ($m_1<m_2<m_3$) in the energy range $M_G-M_Z$. The
initial (GUT) neutrino eigenmasses are all larger than their low energy values.
Significant running is observed mainly for the heaviest eigenmass $m_{\nu_3}$.
For experimentally acceptable mass-squared differences $\Delta m_{\nu_{ij}}^2$ at
$M_Z$, in all cases their corresponding values at the GUT scale lie out of the
acceptable range. In this scenario with hierarchical light neutrino masses, we
find that large renormalization effects occur above the heavy neutrino threshold
since the Yukawa couplings $Y_{\nu}$ are large and the second term in (\ref{mnr})
dominates. Also, since $m_{\nu_1}< \sqrt{\Delta m^2_{sol}}$, the solar angle
turns out to be more stable compared to the running of the $\theta_{23}$, as can
be seen in Figure (\ref{anglesevolatex}). These results are in agreement with the
findings of~\cite{Ellis:2005dr}.

\begin{table}
\begin{center}
\begin{tabular}{|c|c|c|c|c|} \hline Soln. & $\Delta m_{atm}^2(M_G) \cdot 10^{3}$ &
$\Delta m_{sol}^2(M_G)\cdot 10^{5}$ & $\langle m \rangle$ & $b_2$\\
 \hline 1.  & 10.5294 & 27.6096 & 0.00361 &3.41\\
 \hline 2.  & 8.1987 & 30.1333  & 0.00633 &2.88\\
 \hline 3.  & 7.2533 & 31.6108 & 0.00607 &1.50\\
 \hline 4.  & 11.7749 & 30.4352 & 0.00753 &1.28\\
 \hline 5.  & 14.093 & 23.7416 & 0.00346 &2.85\\
 \hline 6.  & 12.347 & 28.2642 & 0.00822 &1.26\\
 \hline 7.  & 7.4256 & 30.855 & 0.00821 &1.23\\
 \hline 8.  & 5.2910 & 29.8775 & 0.00852 &1.17\\
 \hline 9.  & 9.74656 & 30.5418 & 0.00379 &3.64\\
 \hline 10. & 8.99695 & 25.8838 & 0.00327 &3.42\\
\hline
 \end{tabular}
\end{center}
 \caption{Values of the physical parameters $\Delta m^2_{atm}$ and
 $\Delta m^2_{sol}$ at $M_G$; the effective neutrino mass $\vev{m}$
 related to $\beta\beta_{0 \nu}$ decay; and the parameter $b_2$
 which determines the scale of the light matrix $m_{\nu}$.}
 \label{dmthetanumatmgut}
\end{table}

\begin{figure}[tbh]
\centering
\includegraphics[scale=0.70]{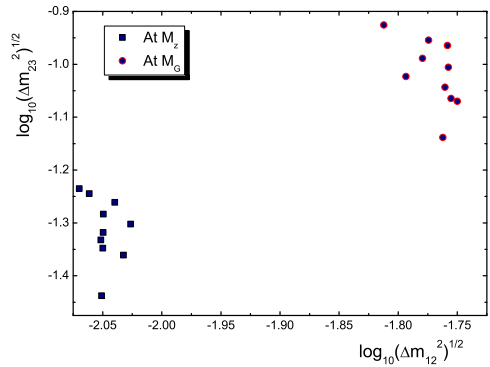}
\caption{$\Delta m^2_{12}$ and $\Delta m^2_{23}$ at $M_Z$ and at
$M_G$. }
 \label{dmmgmz}
\end{figure}

$\bullet$ In figure (\ref{dmmgmz}) we plot the distribution
$\Delta m_{atm}^2$ versus $\Delta m_{sol}^2$ at the two
scales $M_G$ (Table \ref{dmthetanumatmgut}) and $M_Z$ for
the ten models of Table \ref{nummuij}. We find that the hierarchy
of the neutrino masses at the Pati-Salam breaking scale tends
to be greater than that at low energies. Several models predict
$\Delta m_{23}^2/\Delta m_{21}^2$ out of the experimental
range at $M_G$, although after the running at $M_Z$ they are
consistent with the data.

$\bullet$ Finally, we check the predictions of our model
for the effective neutrino mass parameter relevant for
$\beta\beta_{0 \nu}$ decay. This parameter can be written in terms
of the observable quantities as
\begin{multline}
|\langle m \rangle| = \left|(m_1 \cos^2\theta_\odot  + e^{i
\alpha} \sqrt{\Delta m_{sol}} \sin^2 \theta_\odot)
\cos^2\theta_{13}\right. \\
+ \left. \sqrt{\Delta m_{atm}} \sin^2\theta_{13}
 e^{i\beta} \right|.
\label{mef}
\end{multline}

In the last column of Table (\ref{dmthetanumatmgut}) the
$\beta\beta_{0 \nu}$-decay predictions are presented for
solutions 1-10. Many current experiments attempt to measure
this quantity~\cite{Petcov:2005yq}; the best current limit
on the effective mass is given by the Heidelberg--Moscow
collaboration~\cite{heidmo}
\beq \label{eq:current}
\langle m \rangle \le 0.35\, z~ \mbox {eV},
\eeq
where the parameter $z={\cal O}(1)$ allows for uncertainty
arising from nuclear matrix elements.
%physics involved in calculating the $\beta\beta_{0 \nu}$-decay width.

In a recent analysis of neutrinoless double beta
decay~\cite{Choubey:2005rq} the allowable range of the
effective mass parameter was given for specific scenarios.
In the case of the normal hierarchy the bounds are
\beq
0 < \langle m \rangle < 0.007\,\mbox{eV}
\eeq
thus our results are in the experimentally acceptable
region.

\begin{figure}[tbh]
\centering
\includegraphics[scale=0.62]{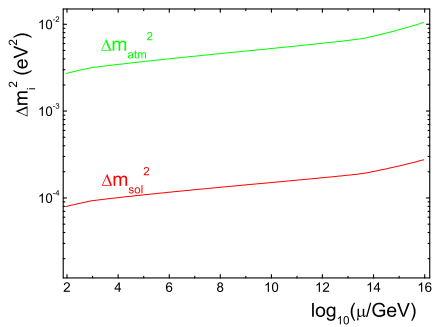}
\caption{Running of $\Delta m^2_{sol}$ and $\Delta m^2_{atm}$.}
 \label{splittingsevolatex}
\end{figure}

%\newpage

\section{Conclusions}

In this work, we studied the running of neutrino masses and mixing
angles in a supersymmetric string-inspired
SU$(4)\times$SU$(2)_L\times $SU$(2)_R\times$U$(1)_X$ model. An
accurate description of the low energy neutrino data forced us to
introduce two singlets charged under the $U(1)_X$, leading to two
expansion parameters. The mass matrices are then constructed
in terms of three expansion parameters
\beq
\eps \equiv \frac{\vev{\phi}}{M_U},\ \
\eps' \equiv \frac{\vev{H\bar{H}}}{M_U^2},\ \
\eps'' \equiv \frac{\vev{\chi}}{M_U},
\eeq
where $\phi$ and $\chi$ are singlets and $H$, $\bar{H}$ the
SU$(4)\times$SU$(2)_R$-breaking Higgses.
% I propose to add:
The model is simplified by
%imposing specific relations between these expansion parameters, and by
the fractional U$(1)_X$ charges of $H$ and $\chi$, which
ensure that the parameter $\eps''$ only appears as a prefactor
to the heavy Majorana neutrino masses.

The expansion parameter arising from the Higgs v.e.v.'s defines
the ratio of the $SU(4)$ breaking scale $M_G$ to the unification
scale $M_U$: we performed a renormalization group analysis
of gauge couplings under this constraint and found successful
unification with the addition of extra states usually present in
a string spectrum.

Assuming that
only the third generation of quarks and charged leptons acquire
masses at tree level and under a specific choice of $U(1)_X$
charges as well as Clebsch factors, we examined the implications
for the light neutrino masses resulting from the see-saw formula.
We found that the light neutrino mass spectrum is hierarchical and
that the mass hierarchy tends to be larger at the GUT scale than
at $M_Z$ due the renormalization group running. The solar mixing
angle $\theta_{12}$ is stable under RG evolution while larger
renormalization effects are found for the atmospheric mixing angle
$\theta_{23}$ and $\theta_{13}$, always with their values at
$M_Z$ in agreement with experiment.

\section*{Acknowledgements}

This research was funded by the program `HERAKLEITOS' of
the Operational Program for Education and Initial Vocational
Training of the Hellenic Ministry of Education under the 3rd
Community Support Framework and the European Social Fund.
TD is supported by the Impuls- und Vernetzungsfond der
Helmholtz-Gesellschaft.

%\newpage

% REFS

 \end{document}